# Modified normalized Rortex/vortex identification method

**Jianming Liu[1,2] and Chaoqun Liu[2*]**

1. School of Mathematics and Statistics, Jiangsu Normal University, Xuzhou 221116, China

2. Department of Mathematics, University of Texas at Arlington, Arlington 76019, USA

**Abstract**

In this paper, a modified normalized Rortex/vortex identification method named $\widetilde{\Omega}_R$ is presented to improve the original $\Omega_R$ method and cure the bulging phenomenon on the iso-surfaces caused by the original $\Omega_R$ method. Detailed mathematical explanations and the relationship with the $Q$ criterion are described. In addition, the new developed formula does not require two original coordinate rotations, and the calculation is greatly simplified. The numerical results are demonstrated to show the effectiveness of the new modified normalized Rortex/vortex identification method.

**Introduction**

Vortex definition and vortex identification methods are still a common concern in the community of fluid mechanics research, attracting many researchers to design various vortex identification methods.[1-4] These vortex identification methods are designed to overcome the deficiencies of the vorticity method. However, these vortex identification methods are not equivalent in themselves,[5] and the physical meanings are not clear. It is also far-fetched to use these quantities to express the intensity of the vortex. Vortex is a rotational motion of the fluid, and the rotation should have a rotating axis. However, most popular vortex identification methods do not provide a clear definition of the axis of rotation. The detailed reviews on these vortex identification methods can be referred to Refs. 6-10. Recently, a new vector called Rortex is proposed to describe the local rotational axis and strength of vortex.[11-12] The Rortex vector represents a rigid rotation extracted from the vorticity. The remainder of the vorticity is antisymmetric shear. The Rortex vector is parallel to the coherent vortex surface, and the iso-surfaces of the magnitude of the Rortex can be used to identify the vortex structures.[6,11-12] Due to the clear physical meaning of the Rortex vector, it has gained a lot of research attentions.[13-21] Although the iso-surfaces of magnitude of the Rortex vector is very effective to capture the coherent vortex structures, the determination of the threshold is still case-dependent.[16-18] Recently, using the idea of the widely used $\Omega$ method,[6,10,17-22] a normalized Rortex/vortex identification method is developed by Dong et al.[23] The new method denoted by $\Omega_R$ has the several important advantages, including: (1) $\Omega_R$ is a dimensionless relative quantity from 0 to 1, which can be used to do statistics and correlation analysis directly; (2) $\Omega_R$ can distinguish the vortex from high vorticity concentration with high shear and exclude boundary layers; (3) $\Omega_R$ is robust to threshold change and can empirically be set as 0.52 to visualize the vortex structures; (4) $\Omega_R$ has the capability of capturing both strong and weak vortices simultaneously. [6,23] Due to these advantages of this method, $\Omega_R$ quickly attracted many people to use. [17,18,21] However the vortex structure shown by this method is not smooth and there is a bulging phenomenon on the iso-surfaces.

---

[*] **Corresponding author:** Chaoqun Liu, Email: cliu@uta.edu

In this paper, a modified method named $\widetilde{\Omega}_R$ is presented to improve the original $\Omega_R$ method and cure the bulging phenomenon on the iso-surfaces. This modified method was established by analyzing the original $\Omega$ method in 2D and 3D flow fields, taking into account the effects of fluid stretching and compression.

In the Rortex method, we believe that the velocity gradient tensor $\nabla \vec{v}$ has complex conjugate eigenvalues $\lambda_{cr} \pm \lambda_{ci} i$ in the region with rotation, and the real unit eigenvector $\vec{r}$ is the direction of the local rotational axis of vortex. Hence, we have

$$\nabla \vec{v} \cdot \vec{r} = \lambda_r \vec{r} \tag{1}$$

where $\lambda_r$ denotes the real eigenvalue and indicates the rate of change on the axis of rotation. In addition, the definition of local rotation axis means there is no cross-velocity gradient on the local rotation axis.[12] In order to obtain the rotation strength, the first coordinate rotation $Q_r$ is used so that the rotation axis is the $Z$ axis.[6] Then

$$\nabla \vec{V} = Q_r \nabla \vec{v} Q_r^T = \begin{bmatrix} \frac{\partial U}{\partial X} & \frac{\partial U}{\partial Y} & 0 \\ \frac{\partial V}{\partial X} & \frac{\partial V}{\partial Y} & 0 \\ \frac{\partial W}{\partial X} & \frac{\partial W}{\partial Y} & \frac{\partial W}{\partial Z} \end{bmatrix} \tag{2}$$

Hence, in the new $XYZ$ frame, the coordinate $Z$ axis is the local rotation axis. From the Rortex vector, the rotation strength is obtained by a second coordinate rotation $P_r$ in the $XY$ plane to make $\left|\frac{\partial U}{\partial Y}\right|$ or $\left|\frac{\partial V}{\partial X}\right|$ take the minimum value. After turning $\theta$ angle, the velocity gradient tensor will become

$$\nabla \vec{V}_\theta = P_r \nabla \vec{V} P_r^T \tag{3}$$

Therefore, the terms in the $2 \times 2$ upper left submatrix of $\nabla \vec{V}_\theta$ are the following

$$\begin{aligned} \left.\frac{\partial U}{\partial Y}\right|_\theta &= \alpha \sin(2\theta + \varphi) - \beta, \\ \left.\frac{\partial V}{\partial X}\right|_\theta &= \alpha \sin(2\theta + \varphi) + \beta, \\ \left.\frac{\partial U}{\partial X}\right|_\theta &= -\alpha \cos(2\theta + \varphi) + \frac{1}{2}\left(\frac{\partial U}{\partial X} + \frac{\partial V}{\partial Y}\right), \\ \left.\frac{\partial V}{\partial Y}\right|_\theta &= \alpha \cos(2\theta + \varphi) + \frac{1}{2}\left(\frac{\partial U}{\partial X} + \frac{\partial V}{\partial Y}\right), \end{aligned} \tag{4}$$

where

$$\alpha = \frac{1}{2}\sqrt{\left(\frac{\partial V}{\partial Y} - \frac{\partial U}{\partial X}\right)^2 + \left(\frac{\partial V}{\partial X} + \frac{\partial U}{\partial Y}\right)^2}, \tag{5}$$

$$\beta = \frac{1}{2}\left(\frac{\partial V}{\partial X} - \frac{\partial U}{\partial Y}\right), \tag{6}$$

The detailed production and the expression of $\varphi$ can be referred to Refs. 6 and 24. In fact, $\beta$ in (6) is the vorticity in the $XYZ$ frame. Then the rotation strength $R = 2(\beta - \alpha)$, and the Rortex vector is defined by

$$\vec{R} = R\vec{r} \qquad (7)$$

Then the original normalized Rortex/vortex identification method $\Omega_R$ is defined as [23]

$$\Omega_R = \frac{\beta^2}{\alpha^2+\beta^2+\epsilon} \qquad (8)$$

where $\epsilon$ is a small parameter to prohibit the computational noise. And $\epsilon$ is proposed as $\epsilon = b_0(\beta^2 - \alpha^2)_{\max}$, where $b_0$ is a small positive number around 0.001~0.002. Although the original method has many advantages, careful observation will reveal that the iso-surfaces formed by $\Omega_R$ are not smooth and have many bulges. Recall the definition of $\Omega$ method,[6,22] using the denotations of the symmetric tensor $\boldsymbol{A} = \frac{1}{2}(\nabla \vec{v} + \nabla \vec{v}^T)$ and the antisymmetric spin tensor $\boldsymbol{B} = \frac{1}{2}(\nabla \vec{v} - \nabla \vec{v}^T)$, then we can formulate $\Omega$ method as

$$\Omega = \frac{b}{a+b+\epsilon} \qquad (9)$$

where $a = \|\boldsymbol{A}\|_F^2$, $b = \|\boldsymbol{B}\|_F^2$, and $\|\cdot\|_F$ is Frobenius norm. For two-dimensional flow, by the Rortex eigenvalue decomposition,[12,24] and Galilean invariance of $\Omega$ method,[25] we can get

$$\nabla \vec{V}_{\theta\min} = \begin{bmatrix} \lambda_{cr} & -(\beta-\alpha) \\ \beta+\alpha & \lambda_{cr} \end{bmatrix} = \begin{bmatrix} \lambda_{cr} & \alpha \\ \alpha & \lambda_{cr} \end{bmatrix} + \begin{bmatrix} 0 & -\beta \\ \beta & 0 \end{bmatrix} \triangleq \boldsymbol{A} + \boldsymbol{B} \qquad (10)$$

and

$$\Omega_{2D} = \frac{b}{a+b+\epsilon} = \frac{\beta^2}{\beta^2+\alpha^2+\lambda_{cr}^2+\epsilon} \qquad (11)$$

Similarly, for three-dimensional flow, [12]

$$\nabla \vec{V}_{\theta\min} = \begin{bmatrix} \lambda_{cr} & -(\beta-\alpha) & 0 \\ \beta+\alpha & \lambda_{cr} & 0 \\ \xi & \eta & \lambda_r \end{bmatrix} = \begin{bmatrix} \lambda_{cr} & \alpha & \frac{1}{2}\xi \\ \alpha & \lambda_{cr} & \frac{1}{2}\eta \\ \frac{1}{2}\xi & \frac{1}{2}\eta & \lambda_r \end{bmatrix} + \begin{bmatrix} 0 & -\beta & -\frac{1}{2}\xi \\ \beta & 0 & -\frac{1}{2}\eta \\ \frac{1}{2}\xi & \frac{1}{2}\eta & 0 \end{bmatrix} \triangleq \boldsymbol{A} + \boldsymbol{B} \qquad (12)$$

and

$$\Omega_{3D} = \frac{\beta^2+\frac{1}{4}(\xi^2+\eta^2)}{\beta^2+\frac{1}{2}(\xi^2+\eta^2)+\alpha^2+\lambda_{cr}^2+\frac{1}{2}\lambda_{cr}^2+\epsilon} \qquad (13)$$

Considering Eqs. (11) and (13), the normalized Rortex/vortex identification method $\Omega_R$ (8) is not completely matched with the $\Omega$ method. In order to reflect the original $\Omega$ method as much as possible, we recommend the following new modified normalized Rortex/vortex identification method

$$\widetilde{\Omega}_R = \frac{\beta^2}{\beta^2+\alpha^2+\lambda_{cr}^2+\frac{1}{2}\lambda_r^2+\epsilon} \qquad (14)$$

And $\widetilde{\Omega}_R$ requires a parameter greater than 0.5 which is same as $\Omega$ or $\Omega_R$ methods. In practice, we also take $\widetilde{\Omega}_R = 0.52$ as a fixed threshold. Here, for the convenience of explanation, $\alpha$ and $\beta$ are used to represent the modified $\widetilde{\Omega}_R$ method. We will then give an explicit formula that does not require the previous $\boldsymbol{Q}_r$ and $\boldsymbol{P}_r$ rotations to get $\alpha$ and $\beta$. It will greatly simplify the solution process.

At present, the $Q$ criterion is a very popular vortex identification method used in engineering.[26] Below we will explain the mathematical relationship between the new modified $\widetilde{\Omega}_R$ method and the popular $Q$ criterion. In fact, neglect the small $\epsilon$

$$\widetilde{\Omega}_R = \frac{\beta^2}{\beta^2+\alpha^2+\lambda_{cr}^2+\frac{1}{2}\lambda_r^2} > 0.5. \tag{15}$$

i.e.

$$\beta^2 - \alpha^2 > \lambda_{cr}^2 + \frac{1}{2}\lambda_r^2 \tag{16}$$

Assume the flow is incompressible, then $\lambda_r = -2\lambda_{cr}$.[5] Hence, we have
$$\beta^2 - \alpha^2 > 3\lambda_{cr}^2 \tag{17}$$
From the matrix (12), $\beta^2 - \alpha^2 = \lambda_{ci}^2$. [12] Therefore, $\lambda_{ci}^2 > 3\lambda_{cr}^2$. That is

$$\left(\frac{\lambda_{cr}}{\lambda_{ci}}\right)^2 < \frac{1}{3} \tag{18}$$

For incompressible flow, the second invariant $Q$ of velocity gradient tensor can be explicitly written as [5]

$$Q = \lambda_{ci}^2 \left(1 - 3\left(\frac{\lambda_{cr}}{\lambda_{ci}}\right)^2\right) \tag{19}$$

$Q > 0$ criterion requires $\left(\frac{\lambda_{cr}}{\lambda_{ci}}\right)^2 < \frac{1}{3}$. Therefore, the mathematical vortex boundaries for $\widetilde{\Omega}_R > 0.5$ and $Q > 0$ are equivalent completely, and all regions described by them are a subset of the region defined by $\lambda_{ci} > 0$. But the new modified $\widetilde{\Omega}_R$ is a dimensionless quantity from 0 to 1. Furthermore, from Eq. (8), $\Omega_R = \frac{\beta^2}{\alpha^2+\beta^2} > 0.5$ and $\beta^2 - \alpha^2 = \lambda_{ci}^2$, we can conclude that the mathematical vortex boundaries defined by $\Omega_R > 0.5$ and $\lambda_{ci} > 0$ are same. Although the region defined by $\widetilde{\Omega}_R > 0.5$ is a subset defined by $\Omega_R > 0.5$, Eq. (18) will avoid regions of strong outward spiraling given by $\left(\frac{\lambda_{cr}}{\lambda_{ci}}\right)^2 > \frac{1}{3}$. [5] Furthermore, from our computation, the new modified $\widetilde{\Omega}_R$ method can keep smooth and get rid of the bulges on the iso-surfaces.

Recently, Wang et al gave an explicit formula of Rortex vector to simplify the calculations of $\beta$ and $\alpha$, which can be reformulated as

$$R = \vec{\omega}\cdot\vec{r} - \sqrt{(\vec{\omega}\cdot\vec{r})^2 - 4\lambda_{ci}^2} \tag{20}$$

where $\vec{\omega}$ represents the vorticity and $\vec{\omega}\cdot\vec{r}$ is always set to be positive. [27] Then we can obtain

$$\beta = \frac{1}{2}\vec{\omega}\cdot\vec{r} \tag{21}$$

$$\alpha = \frac{1}{2}\sqrt{(\vec{\omega}\cdot\vec{r})^2 - 4\lambda_{ci}^2} \tag{22}$$

Hence, the present new modified normalized Rortex/vortex identification method (14) can be rewritten as

$$\widetilde{\Omega}_R = \frac{(\vec{\omega}\cdot\vec{r})^2}{2[(\vec{\omega}\cdot\vec{r})^2-2\lambda_{ci}^2+2\lambda_{cr}^2+\lambda_r^2]+\epsilon} \tag{23}$$

The new formula (23) does not require the coordinate rotations $Q_r$ and $P_r$, and the calculation is greatly simplified.

In order to show that the modified normalized Rortex/vortex identification method (14) or (23) can get rid of bulging phenomenon performed by original $\Omega_R$ method, the data obtained by direct numerical simulation (DNS) of boundary layer transition on a flat plate at Mach number 0.5 and Reynolds number 1000 are studied. Fig. 1 shows the iso-surfaces obtained by the original and modified normalized Rortex vortex identification methods. By the original $\Omega_R$ method (8), the iso-surfaces lost smoothness and show the emergence of the bulging phenomenon as depicted in Fig. 1(a). The result obtained by the new modified $\widetilde{\Omega}_R$ method is shown in Fig. 1(b), which demonstrates the effectiveness of the new modified method. $\widetilde{\Omega}_R$ is a dimensionless relative quantity from 0 to 1. The iso-surfaces formed by $\widetilde{\Omega}_R$ method is not sensitive to the threshold change. For different examples, it can always take $\widetilde{\Omega}_R = 0.52$, and it has the shape-preserving feature as the threshold increases. However, for the popular $Q$ criterion, the determination of the proper threshold is unclear in advance and needs to be determined case by case, sometimes could be as large as $10^8$ (see Ref. 18), but the current example only takes a small value. Therefore, Q is very sensitive to the threshold selection while $\widetilde{\Omega}_R$ is not. As a demonstration, we compare the results of the iso-surfaces obtained by $Q$ criterion and the new modified $\widetilde{\Omega}_R$. The results obtained by $Q$ criterion at $Q = 0.005, 0.02, 0.04, 0.05$ are shown in Figs. 2a, 2c, 2e, and 2g. The vortex structures obtained by $\widetilde{\Omega}_R$ method at $\widetilde{\Omega}_R = 0.52, 0.6, 0.7, 0.8$ are shown in Figs. 2b, 2d, 2f, and 2h. The graph of the DNS data shows that as the threshold of the $Q$ criterion increases, many vortex structures disappear in the downstream of the flow, especially the ring structure. The $\widetilde{\Omega}_R$ method still maintains the vortex structures as the threshold increases. Furthermore, the vortex intensity represented by the $Q$ has been greatly reduced in downstream, but the relative vortex strength is still relatively large as shown by the $\widetilde{\Omega}_R$ method.

In this paper, a modified normalized Rortex/vortex identification method named $\widetilde{\Omega}_R$ is presented to improve the original $\Omega_R$ method and cure the bulging phenomenon on the iso-surfaces. Detailed mathematical explanations are provided. Also the relationship with the $Q$ criterion is described. The developed new formula does not require original coordinate rotations $\boldsymbol{Q}_r$ and $\boldsymbol{P}_r$, and the calculation is greatly simplified. Furthermore, the developed new modified $\widetilde{\Omega}_R$ method retain the advantages of original $\Omega_R$ method. It is a dimensionless relative quantity from 0 to 1, which can be used to do statistics and correlation analysis directly. $\widetilde{\Omega}_R$ can distinguish vortex from high vorticity concentration with high shear and exclude boundary layers. $\widetilde{\Omega}_R$ is robust and can always be set as 0.52 to visualize the vortex structures. $\widetilde{\Omega}_R$ has the capability of capturing both strong and weak vortices simultaneously. In this paper, we also compared the results with $Q$ criterion. The results show the $\widetilde{\Omega}_R$ method can retain the vortex ring structures when the threshold increases. In many places, the relative strength of vortex structures is still high.

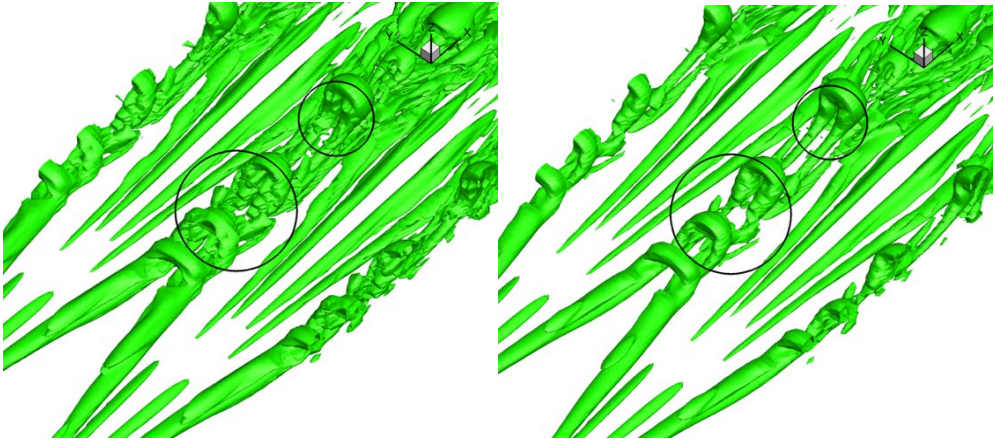

(a) Old $\Omega_R$ method (many bulges)   (b) New $\widetilde{\Omega}_R$ method (smooth)

FIG. 1 Iso-surfaces of hairpin vortex structures

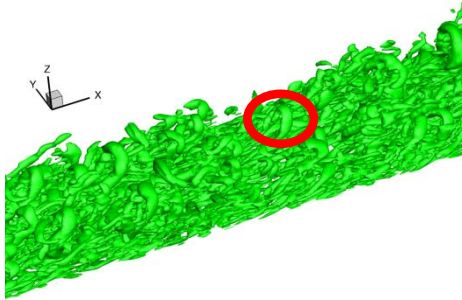

(a) $Q = 0.005$

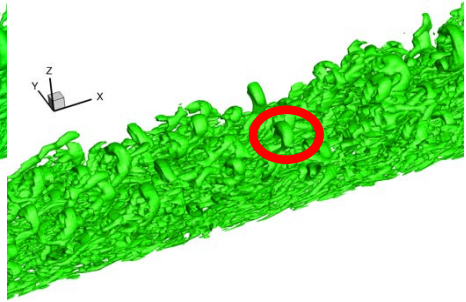

(b) $\widetilde{\Omega}_R = 0.52$

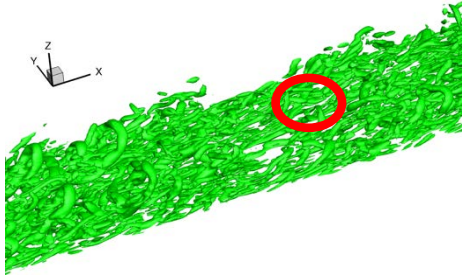

(c) $Q = 0.02$

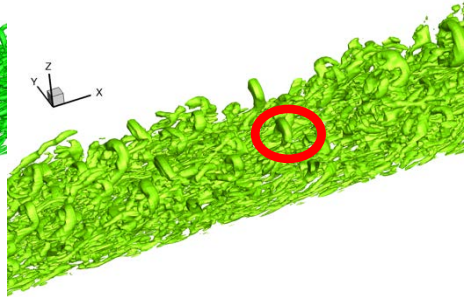

(d) $\widetilde{\Omega}_R = 0.6$

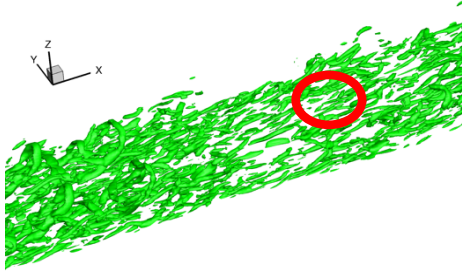

(e) $Q = 0.04$

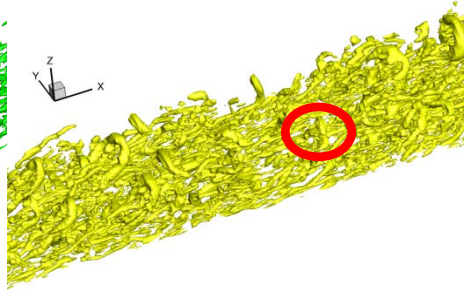

(f) $\widetilde{\Omega}_R = 0.7$

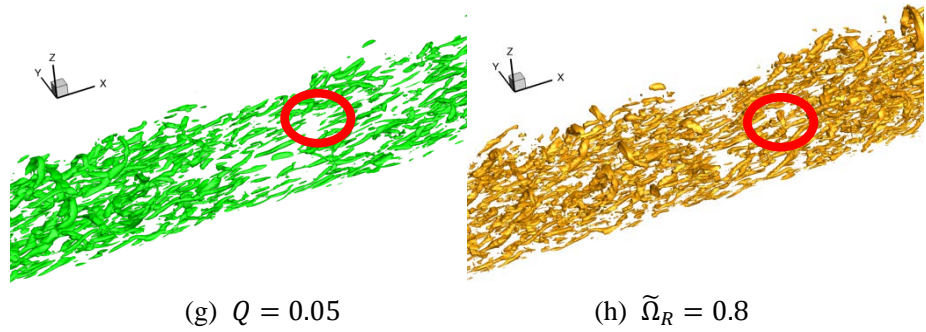

(g) $Q = 0.05$  (h) $\widetilde{\Omega}_R = 0.8$

FIG. 2 Iso-surfaces vortex structures obtained by $Q$ criterion (rings disappear) and $\widetilde{\Omega}_R$ method (ring's structure remains)

**Acknowledgments**

This work was mainly supported by the Department of Mathematics of University of Texas at Arlington. The research was partly supported by the National Natural Science Foundation of China (Grant No. 91530325) and the Natural Science Foundation of the Jiangsu Higher Education Institutions of China (Grant Nos.18KJA110001), the Visiting Scholar Scholarship of the China Scholarship Council (Grant No. 201808320079). The authors are grateful to Texas Advanced Computational Center (TACC) for providing computation hours. This work is accomplished by using code DNSUTA developed by Dr. Chaoqun Liu at the University of Texas at Arlington.